\documentclass{elsart}

\usepackage{graphicx}

\usepackage{amssymb}

\usepackage{amsmath}

\DeclareMathOperator{\tr}{tr}

\begin{document}

\begin{frontmatter}



\title{Some physics of the two-dimensional $\mathcal{N}=(2,2)$ supersymmetric
Yang-Mills theory:\\
Lattice Monte Carlo study}


\author{Issaku Kanamori},
\ead{kanamori-i@riken.jp}
\author{Hiroshi Suzuki}
\ead{hsuzuki@riken.jp}
\address{Theoretical Physics Laboratory, RIKEN, Wako 2-1, Saitama 351-0198,
Japan}

\begin{abstract}
We illustrate some physical application of a lattice formulation of the
two-dimensional $\mathcal{N}=(2,2)$ supersymmetric $SU(2)$ Yang-Mills theory
with a (small) supersymmetry breaking scalar mass. Two aspects, power-like
behavior of certain correlation functions (which implies the absence of the
mass gap) and the static potential~$V(R)$ between probe charges in the
fundamental representation, are considered. For the latter, for~$R\lesssim1/g$,
we observe a linear confining potential with a finite string tension. This
confining behavior appears distinct from a theoretical conjecture that a probe
charge in the fundamental representation is screened in two-dimensional gauge
theory with an adjoint massless fermion, although the static potential
for~$R\gtrsim1/g$ has to be systematically explored to conclude real asymptotic
behavior in large distance.
\end{abstract}

\end{frontmatter}

\section{Introduction}
\label{sec:1}
Recently, through the observation of a ``partially conserved supercurrent
relation'', we obtained~\cite{Kanamori:2008bk} an affirmative numerical
evidence that a lattice formulation in~Ref.~\cite{Sugino:2004qd} provides a
supersymmetric regularization of the two-dimensional $\mathcal{N}=(2,2)$
supersymmetric Yang-Mills theory (SYM)
\footnote{For other lattice formulations of this system, see
Refs.~\cite{Kaplan:2002wv,Cohen:2003xe,Sugino:2003yb,Catterall:2004np,%
Suzuki:2005dx,D'Adda:2005zk,Sugino:2006uf}. For recent developments in this
field of research, see Ref.~\cite{Giedt:2007hz} for a review and references
cited in~Ref.~\cite{Kanamori:2008bk}.
As further recent study, see
Refs.~\cite{Unsal:2008kx,Demmouche:2008ms,Endres:2008tz,Ishiki:2008te,%
Giedt:2008xm,Kikukawa:2008xw,Catterall:2008dv,Demmouche:2008aq,Hanada:2008gy}.}%
\footnote{This system can be obtained by dimensionally reducing the
four-dimensional $\mathcal{N}=1$ SYM from four to two dimensions and hence a
four-dimensional notation is useful; Roman indices~$M$ and $N$ run over 0, 1,
2 and~3, while Greek indices~$\mu$ and~$\nu$ below run over only 0 and~1. With
the dimensional reduction, it is understood that $\partial_2=0$
and~$\partial_3=0$. $\Psi$ is a four-component spinor. We follow the notational
convention in~Ref.~\cite{Kanamori:2008bk}. Note that the gauge coupling~$g$
has the mass dimension~1.}
\begin{equation}
   S=\frac{1}{g^2}
   \int d^2x\,\tr\left\{
   \frac{1}{2}F_{MN}F_{MN}+\Psi^TC\Gamma_MD_M\Psi+\widetilde H^2\right\},
\label{one}
\end{equation}
when one supplements to $S$ a supersymmetry breaking scalar mass term
\begin{equation}
   S_{\text{mass}}=\frac{1}{g^2}\int d^2x\,\mu^2
   \tr\left\{A_2A_2+A_3A_3\right\}.
\label{two}
\end{equation}
The scalar mass term was added to suppress a possible large amplitude of scalar
fields along flat directions that may amplify $O(a)$~lattice artifacts
to~$O(1)$~\cite{Kanamori:2008bk}. In the present Letter, we illustrate some
physical application of this lattice formulation for the
system~$S+S_{\text{mass}}$.

\section{Correlation functions with power-like behavior}
\label{sec:2}

Assuming the 't~Hooft anomaly matching condition, in Ref.~\cite{Witten:1995im},
it was pointed out that the two-dimensional $\mathcal{N}=(2,2)$ SYM has no mass
gap. This aspect has been numerically investigated from almost a decade
ago~\cite{Antonuccio:1998mq,Harada:2004ck} by utilizing the
supersymmetric discretized light-cone formulation~\cite{Matsumura:1995kw}. In
this super-renormalizable system, it is in fact possible to determine (to all
orders of perturbation theory) an explicit form of a correlation function
between Noether currents, by employing anomalous Ward-Takahashi (WT) identities
(i.e. the Kac-Moody algebra)~\cite{Fukaya:2006mg}; this explicit form directly
proves the above assertion. Here, rather than supersymmetry, continuous global
(bosonic) symmetries are important and the proof~\cite{Fukaya:2006mg} applies
even with supersymmetry breaking scalar mass term~(\ref{two}).

The total action~$S+S_{\text{mass}}$ is invariant under the (two-dimensional)
$U(1)_V$ transformation, $\Psi\to\exp\{i\alpha\Gamma_5\}\Psi$, and an
associated Noether current ($U(1)_V$ current) is given by
\begin{equation}
   j_\mu\equiv
   \frac{1}{g^2}\tr\left\{\Psi^TC\Gamma_\mu\Gamma_5\Psi\right\}.
\label{three}
\end{equation}
Similarly, associated with the $U(1)_A$ symmetry,
$\Psi\to\exp\{\alpha\Gamma_2\Gamma_3\}\Psi$,
$A_2\to\cos\{2\alpha\}A_2-\sin\{2\alpha\}A_3$ and
$A_3\to\sin\{2\alpha\}A_2+\cos\{2\alpha\}A_3$, there is a Noether current
($U(1)_A$ current),
\begin{equation}
   j_{5\mu}\equiv
   \frac{1}{g^2}\tr\left\{-i\Psi^TC\Gamma_\mu\Gamma_2\Gamma_3\Psi
   +4i(A_3F_{\mu2}-A_2F_{\mu3})\right\}.
\label{four}
\end{equation}
It is then possible to show that~\cite{Fukaya:2006mg}, for the two-dimensional
euclidean space~$\mathbb{R}^2$,
\begin{align}
   &-\frac{i}{2}\left\langle j_\mu(x)\epsilon_{\nu\rho}j_{5\rho}(0)\right\rangle
\nonumber\\
   &=\frac{1}{4\pi}(N_c^2-1)\int\frac{d^2p}{(2\pi)^2}\,e^{ipx}\left\{
   -\frac{1}{p^2}(p_\mu p_\nu-\epsilon_{\mu\rho}\epsilon_{\nu\sigma}p_\rho p_\sigma)
   +\widetilde c\delta_{\mu\nu}\right\}
\nonumber\\
   &=\frac{1}{4\pi}(N_c^2-1)
   \left\{\frac{1}{\pi}\frac{1}{(x^2)^2}
   (x_\mu x_\nu-\epsilon_{\mu\rho}\epsilon_{\nu\sigma}x_\rho x_\sigma)
   +\widetilde c\delta_{\mu\nu}\delta^2(x)\right\},
\label{five}
\end{align}
to all orders of perturbation theory, where $N_c$ is the number of colors and
the constant~$\widetilde c$ is a regularization ambiguity in a divergent
one-loop diagram. Thus the correlation function between the $U(1)_V$ current
and the $U(1)_A$ current possesses a massless pole and this is precisely what
the 't~Hooft anomaly matching condition claims for this two-dimensional system.

We want to confirm the power-like behavior of correlation function
in~Eq.~(\ref{five}) by using a lattice Monte Carlo simulation. For this, we
prepared sets of uncorrelated configurations listed in~Table~\ref{table:1}.
For simulation details, see
Refs.~\cite{Kanamori:2008vi,Kanamori:2008ve,Kanamori:2008bk}. In the table, $a$
denotes the lattice spacing and $\beta$ and $L$ are temporal and spatial
physical sizes of our lattice, respectively. The scalar mass squared is
$\mu^2/g^2=0.25$ for all cases. The temporal boundary condition for fermionic
variables is antiperiodic as in~Ref.~\cite{Kanamori:2008bk}. For current
operator~(\ref{four}), we discretized the covariant derivatives
$F_{\mu2}=D_\mu A_2$ and~$F_{\mu3}=D_\mu A_3$ by using the forward covariant
lattice difference. Eq.~(\ref{five}) suggests that we should not take an
average of the correlation function over the spatial coordinate~$x_1$ (i.e.,
projection to the zero spatial momentum) because after the average, correlation
function~(\ref{five}) becomes proportional to $\delta(x_0)$ that cannot be
distinguished from the regularization ambiguity; we should measure the
correlation function as it stands without the zero spatial momentum projection.
\begin{table}
\caption{Sets of uncorrelated configurations used for Figs.~\ref{fig:1}
and~\ref{fig:2}. The scalar mass squared is $\mu^2/g^2=0.25$ for all cases.}
\label{table:1}
\begin{center}
\begin{tabular}{ccccc}
\hline\hline
lattice size & $ag$ & $\beta g\times Lg$ & number of configurations
& set label\\
\hline
$16\times8$ & 0.1768 & $2.828\times1.414$ & 400 & I \\
$20\times10$ & 0.1414 & $2.828\times1.414$ & 800 & II \\
$24\times12$ & 0.1179 & $2.828\times1.414$ & 400 & III \\
$20\times16$ & 0.1414 & $2.828\times2.263$ & 400 & IV \\
\hline\hline
\end{tabular}
\end{center}
\end{table}

\begin{figure}[htbp]
\begin{center}
\includegraphics*[width=\textwidth,clip]{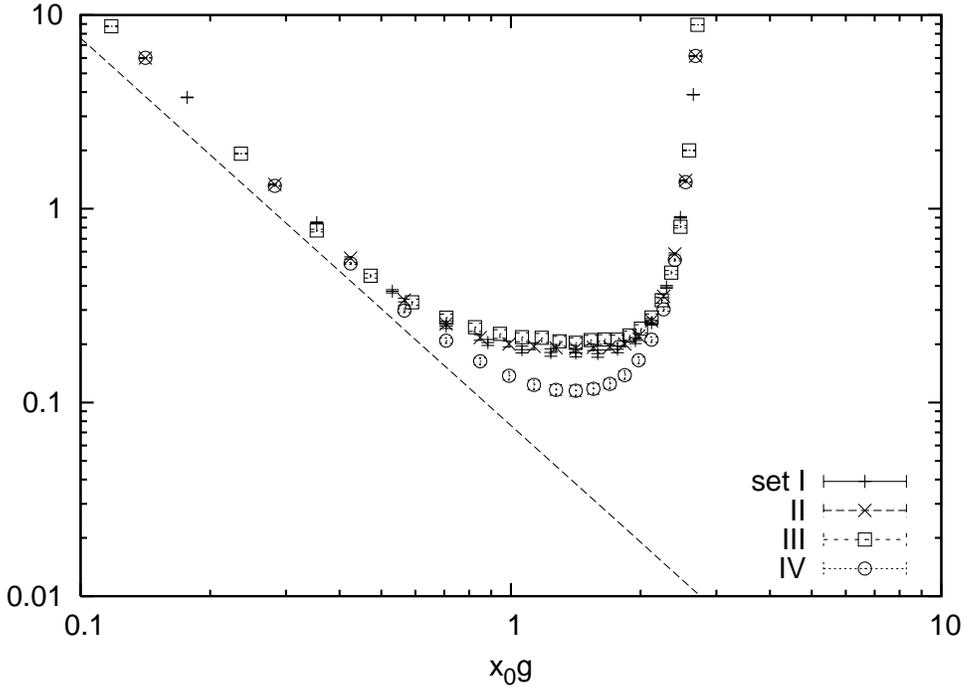}
\end{center}
\caption{The correlation function
$-i\langle j_\mu(x)\epsilon_{\nu\rho}j_{5\rho}(0)\rangle/(2g^2)$ with
$\mu=\nu=0$ along the line~$x_1=0$, for the configuration sets
in~Table~\ref{table:1}. The broken line is theoretical prediction~(\ref{five})
for $\mathbb{R}^2$.}
\label{fig:1}
\end{figure}

\begin{figure}[htbp]
\begin{center}
\includegraphics*[width=\textwidth,clip]{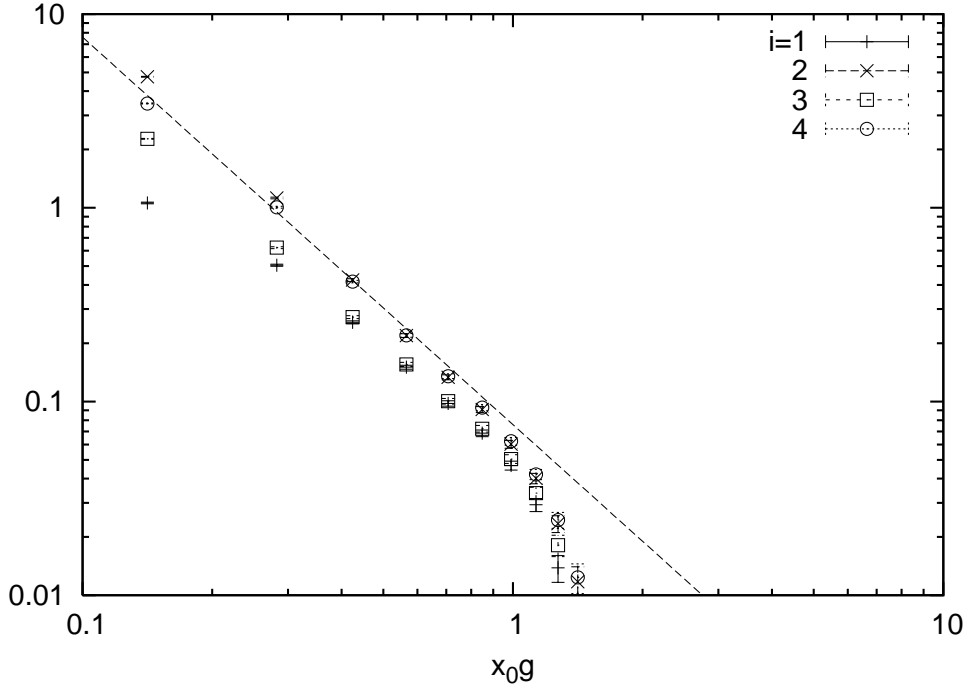}
\end{center}
\caption{The correlation function
$\langle(s_\mu)_i(x)(f_\nu)_i(0)\rangle/g^2$ with $\mu=\nu=0$ along the
line~$x_1=0$ for set~IV in~Table~\ref{table:1}. The broken line is
$(3/4\pi^2)1/(x_0)^2$, the same function plotted in~Fig.~\ref{fig:1}.}
\label{fig:2}
\end{figure}

In~Fig.~\ref{fig:1}, we plotted
$-i\langle j_\mu(x)\epsilon_{\nu\rho}j_{5\rho}(0)\rangle/2$ with
$\mu=\nu=0$ along the line~$x_1=0$. We plotted also theoretical
prediction~(\ref{five}) for $\mathbb{R}^2$ with $N_c=2$, $(3/4\pi^2)1/(x_0)^2$,
by the broken line.
We clearly see the power-like fall of the correlation function
for~$x_0g\lesssim0.7$\footnote{In Fig.~\ref{fig:1}, we plotted the correlation
function as a function of $x_0$, along the line~$x_1=0$. As $x_0$ moves away
from the origin $x_0=0$, the point $x$ approaches a periodic image of the
origin at~$x_0=\beta$ and for $x_0\gtrsim\beta/2$ we expect the correlation
function is power-like in the variable $\beta-x_0$. In other words, the fact
that our finite-size lattice is topologically~$T^2$ but not~$\mathbb{R}^2$
cannot be neglected for $x_0\gtrsim\beta/2$. We thus do not expect the
power-like fall (that is expected for ${\mathbb R}^2$) for~$x_0g\gtrsim1$ and
actually the plot blows up for $x_0g\gtrsim1$ (in our simulation,
$\beta g=2.828$). This remark is applied also to Fig.~\ref{fig:2}, in which the
antiperiodic boundary condition for fermionic fields implies ``blow-down'' for
$x_0g\gtrsim1$.}
instead of exponential one, although the overall amplitude is somewhat larger
than the theoretical expectation for~$\mathbb{R}^2$. From the behavior in the
figure, we think that this discrepancy in the overall amplitude is caused by a
finite lattice spacing and volume. In particular, comparison between set~II
(indicated by~$\times$) and set~IV (indicated by~$\bigcirc$) shows that the
finite size effect is rather large (note that these two sets differ only in the
spatial physical size~$L$). We thus expect that the theoretical prediction
for~$\mathbb{R}^2$ is eventually reproduced in the limit, $a\to0$ and $\beta$,
$L\to\infty$, although we do not carry out a systematic study on this limit.

What is the implication of the above observation? It indicates that our target
theory, the two-dimensional $\mathcal{N}=(2,2)$ $SU(2)$ SYM with a scalar mass
term, is realized in the continuum limit of the present lattice model. In
particular, in deriving Eq.~(\ref{five}), one assumes that the $U(1)_V$
and~$U(1)_A$ currents $j_\mu$ and~$j_{5\nu}$ individually
conserve~\cite{Fukaya:2006mg}.\footnote{This assumption fails for example in
the massless Schwinger model, in which the $U(1)_A$ current suffers from the
axial anomaly; note that the massless Schwinger model has a mass gap.}
One assumes $U(1)_V$ and~$U(1)_A$ symmetries in this sense. In the present
lattice formulation~\cite{Sugino:2004qd}, the $U(1)_V$ symmetry is explicitly
broken for finite lattice spacings. The above observation hence indicates that
the $U(1)_V$ symmetry is fairly restored with present lattice spacings. (This
symmetry will eventually be restored in the continuum
limit~\cite{Kanamori:2008bk}.)

Now, if the system were supersymmetric, and if supersymmetry is not
spontaneously broken, there would exist a massless fermionic state
corresponding to the massless bosonic state appearing in~Eq.~(\ref{five})
as an intermediate state. We expect that this fermionic state produces a
massless pole in the correlation functions
\begin{equation}
   \left\langle(s_\mu)_i(x)(f_\nu)_i(0)\right\rangle\qquad
   \text{($i=1$, 2, 3, 4; no sum over $i$)},
\label{six}
\end{equation}
where $i$ denotes the spinor index and
\begin{align}
   &s_\mu\equiv
   -\frac{1}{g^2}C\Gamma_M\Gamma_N\Gamma_\mu\tr\left\{F_{MN}\Psi\right\},
\label{seven}\\
   &f_\mu\equiv
   \frac{1}{g^2}\Gamma_\mu\left(
   \Gamma_2\tr\{A_2\Psi\}+\Gamma_3\tr\{A_3\Psi\}\right).
\label{eight}
\end{align}
In the above, $s_\mu$ is the supercurrent associated with the supersymmetry
of~$S$, $\delta A_M=i\epsilon^TC\Gamma_M\Psi$,
$\delta\Psi=\frac{i}{2}F_{MN}\Gamma_M\Gamma_N\epsilon
+i\widetilde H\Gamma_5\epsilon$,
and $\delta\widetilde H=-i\epsilon^TC\Gamma_5\Gamma_MD_M\Psi$,
and $f_\mu$ is a lowest-dimensional fermionic spinor-vector (considered
in~Ref.~\cite{Kanamori:2008bk}). Eq.~(\ref{six}) with $i=1$, 2, 3, and~4 are
precisely four correlation functions studied in~Eq.~(11)
of~Ref.~\cite{Kanamori:2008bk} and, as noted there, these four functions are
identical to each other in the continuum theory. Our expectation that
Eq.~(\ref{six}) possesses a massless pole stems from a supersymmetric WT
identity for a vanishing scalar mass squared, $\mu^2=0$
\begin{align}
   &\frac{1}{4}\sum_{i=1}^4
   \left\langle(s_\mu)_i(x)(f_\nu)_i(0)\right\rangle
\nonumber\\
   &=-\frac{i}{2}\left\langle j_\mu(x)\epsilon_{\nu\rho}j_{5\rho}(0)\right\rangle
   -\left\langle j_\mu(x)\epsilon_{\nu\rho}\frac{1}{g^2}
   \tr\left\{A_3(0)F_{\rho2}(0)-A_2(0)F_{\rho3}(0)\right\}\right\rangle,
\label{nine}
\end{align}
which follows from $\delta\langle j_\mu(x)f_\nu^T(0)\rangle=0$, where $\delta$
is the \emph{global\/} super transformation; this relation holds under the
assumptions that the boundary condition is consistent with supersymmetry and
supersymmetry is not spontaneously broken. (In deriving Eq.~(\ref{nine}), we
have used also the equation of motion of the auxiliary field,
$\widetilde H\equiv H-iF_{01}=0$). In the right-hand side of~Eq.~(\ref{nine}),
the massless pole in the first term (recall Eq.~(\ref{five})) cannot be
cancelled by the second term, because the latter is~$O(g^2)$ as one can easily
see.\footnote{In fact, a one-loop calculation of Eq.~(\ref{six}) on the basis
of Eq.~(\ref{one}) coincides with expression~(\ref{five}), possibly with a
different regularization-dependent constant~$\tilde c$. Note also that even if
correlation functions~(\ref{six}) possess a massless pole of
structure~(\ref{five}), this does \emph{not\/} necessarily imply the existence
of the Nambu-Goldstone fermion associated with the spontaneous supersymmetry
breaking. This is because structure~(\ref{five}) shows that the Fourier
transform of $\partial_\mu\langle(s_\mu)_i(x)(f_\nu)_i(0)\rangle$ vanishes at
zero momentum.}

Even if the supersymmetry breaking owing to lattice regularization disappears
in the continuum limit~\cite{Kanamori:2008bk}, our present system is not
supersymmetric because there is scalar mass term~(\ref{two}) and we used the
antiperiodic temporal boundary condition for fermions. These will give
additional contribution to~Eq.~(\ref{nine}). In~Fig.~\ref{fig:2}, we plotted
correlation functions~(\ref{six}) along the line $x_1=0$ for set~IV
in~Table~\ref{table:1}. (For the parameters of this configuration set,
naively-expected order of magnitude of supersymmetry breaking caused by above
factors would be $\sim\mu=0.5g$ and~$\sim1/\beta\simeq0.3536g$, respectively.)
For $0.2\lesssim x_0g\lesssim1.0$, for all~$i$, the power-like behavior
expected from supersymmetric WT identity~(\ref{nine}) combined
with~Eq.~(\ref{five}) is fairly observed. Somewhat surprisingly, we do not see
a significant effect of the supersymmetry breaking and it appears that the
fermionic intermediate state is approximately massless as expected from
approximate supersymmetry. This result is consistent with the conclusion
of~Ref.~\cite{Kanamori:2008bk} that the supersymmetry breaking owing to the
lattice regularization disappears in the continuum limit.

\section{Potential energy between probe charges in the
fundamental representation}
\label{sec:3}
Contrary to naive intuition, it is believed that a probe charge in the
\emph{fundamental\/} representation is screened by dynamical \emph{adjoint\/}
massless fermions in the two-dimensional $SU(N_c)$
QCD~\cite{Gross:1995bp,Armoni:1997ki}. This phenomenon is analogous to the
screening of a fractional charge in the massless Schwinger model with an
integer-charged fermion and is believed to occur also in the two-dimensional
$\mathcal{N}=(1,1)$ SYM, despite the presence of a scalar field and a Yukawa
interaction in the latter~\cite{Gross:1995bp,Armoni:1998kv} (see also
Ref.~\cite{Armoni:1999xw}). As a generalization of these,
in~Refs.~\cite{Armoni:1998kv,Armoni:1999xw}, it was claimed that this screening
persists in any two-dimensional (supersymmetric and non-supersymmetric) gauge
theory with adjoint massless fermions, although an explicit proof was not given
there. In our present system, the masslessness of the gaugino is ensured by
the global $U(1)_A$ and~$U(1)_V$ symmetries and, hence, it is of interest to
study the static potential energy between probe charges in the fundamental
representation. If the expected screening occurs, the static potential would
approach a constant for large distance (i.e., the Wilson loop obeys the
perimeter law).

\begin{figure}[htbp]
\begin{center}
\includegraphics*[width=\textwidth,clip]{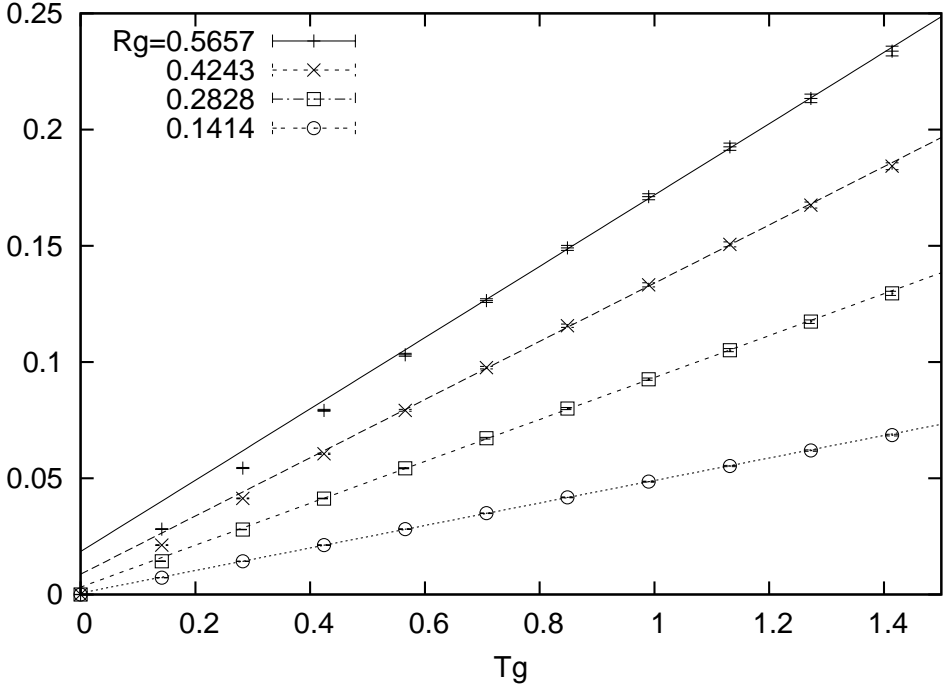}
\end{center}
\caption{The linear $\chi^2$-fit of ${-\ln\{W(T,R)\}}$ for set~VI
in~Table~\ref{table:2}.}
\label{fig:3}
\end{figure}

We thus measure the expectation value of the Wilson loop,
\begin{equation}
   W(T,R)\equiv
   \left\langle\frac{1}{2}\tr\left\{\prod_{\ell\in C}U_\ell\right\}\right\rangle,
\end{equation}
where $C$ denotes a rectangular loop of a physical size~$T\times R$ and link
variables~$U_\ell$ belong to the fundamental representation of the gauge
group~$SU(2)$. For this average, we prepared uncorrelated configurations
listed in~Table~\ref{table:2} (the scalar mass squared is $\mu^2/g^2=0.25$ for
all cases).
\begin{table}
\caption{Sets of uncorrelated configurations used for Figs.~\ref{fig:3}
and~\ref{fig:4}. The scalar mass squared is~$\mu^2/g^2=0.25$ for all cases.}
\label{table:2}
\begin{center}
\begin{tabular}{ccccc}
\hline\hline
lattice size & $ag$ & $\beta g\times Lg$ & number of configurations
& set label\\
\hline
$20\times10$ & 0.2 & $4\times2$ & 800 & V \\
$20\times10$ & 0.1414 & $2.828\times1.414$ & 800 & VI \\
$20\times16$ & 0.1414 & $2.828\times2.263$ & 400 & VII \\
$30\times10$ & 0.1414 & $4.243\times1.414$ & 800 & VIII \\
\hline\hline
\end{tabular}
\end{center}
\end{table}
Theoretically, the static potential $V(R)$ is defined by the asymptotic form
in~$T\to\infty$:
\begin{equation}
   -\ln\left\{W(T,R)\right\}=V(R)T+c(R).
\label{eleven}
\end{equation}
Practically, with finite-size lattices, we made a linear $\chi^2$-fit
of~${-\ln\{W(T,R)\}}$ with respect to~$T$ in a finite range
$T_{\text{min}}\leq T\leq\beta/2$ for each~$R$ and regarded the slope as~$V(R)$;
obviously $\beta/2$ is a maximal temporal size of the Wilson loop that is
physically meaningful. We determined the lower end of the fit~$T_{\text{min}}$
such that the fitting range becomes as wide as possible insofar as
$\chi^2/\text{dof}$ of the fit does not exceed unity. We had $T_{\text{min}}=a$
-- $5a$. A typical result of this linear fit is depicted in~Fig.~\ref{fig:3}
for the case of set~VI in~Table~\ref{table:2}.

\begin{figure}[htbp]
\begin{center}
\includegraphics*[width=\textwidth,clip]{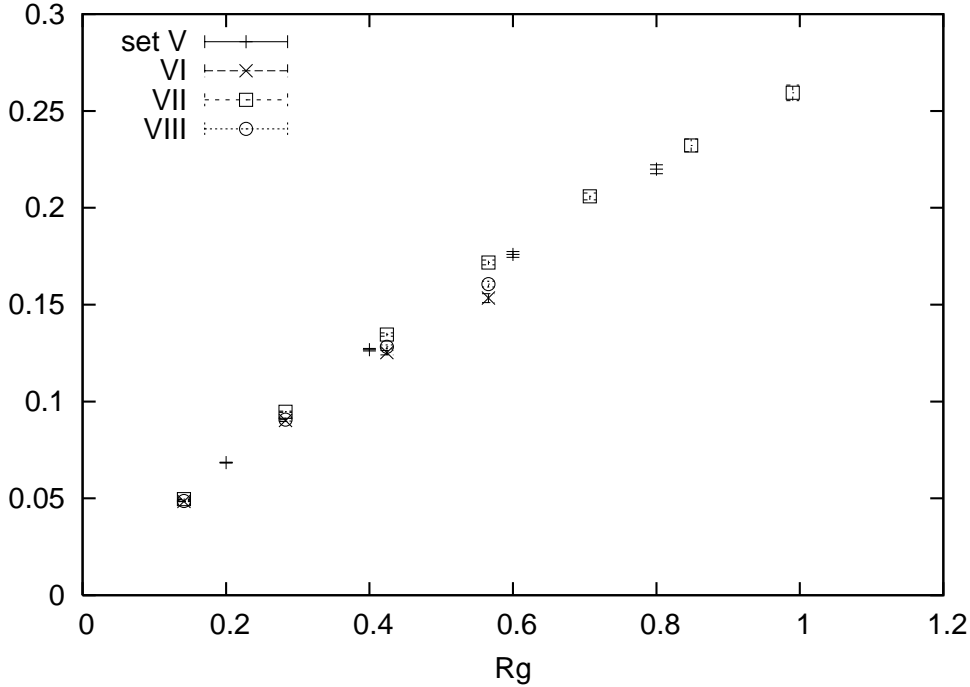}
\end{center}
\caption{$V(R)/g$ determined by the linear $\chi^2$-fit described in the text.
See~Table~\ref{table:2} for the label of configuration sets.
$\mu^2/g^2=0.25$.}
\label{fig:4}
\end{figure}

In~Fig.~\ref{fig:4}, we plotted $V(R)$ for $R<L/2$ determined in this way
for various lattice spacings and lattice sizes (Table~\ref{table:2}). The error
in the figure was determined by the range of a slope of the linear fit that
corresponds to a unit variation of~$\chi^2$.

Now in Fig.~\ref{fig:4}, all points are almost on a common line, although
lattice spacings are different ($ag=0.2$ for~$+$ and~$ag=0.1414$ for~$\times$,
$\square$ and~$\bigcirc$). This fact indicates that the result
in~Fig.~\ref{fig:4} can roughly be regarded as that in the continuum limit.
Similarly, since physical lattice sizes of each configuration set are rather
different (for example, $Lg=1.414$ for $\times$ and $Lg=2.263$ for~$\square$),
there appears almost no significant finite-size effect.\footnote{The
discrepancy between $\square$ and~$\times$ at $Rg=0.5657$ could be explained by
the fact that this value of~$Rg$ is comparable with the spatial lattice size
for~set~VI. Note that for smaller $Rg$ they have less discrepancies.}
Therefore, at least for~$Rg\lesssim1$, we could conclude that the static
potential~$V(R)$ is linear (i.e., the Coulomb potential in two dimensions) with
a finite string tension~$\sigma\sim0.25g^2$. This appears to be distinct from a
theoretical conjecture in~Refs.~\cite{Armoni:1998kv,Armoni:1999xw}.\footnote{A
possible confutation is that the gaugino is not strictly massless in our
simulation because of the antiperiodic temporal boundary condition. This point
seems irrelevant, however, because the behavior in Fig.~\ref{fig:4} appears
insensitive to the temporal size~$\beta$ of our lattice. Compare, for example,
set~VI and set~VIII.} However, to conclude whether a probe charge is really
confined or screened, the static potential for~$Rg\gtrsim1$ has to be
systematically explored; we reserve this as a future project.

\begin{figure}[htbp]
\begin{center}
\includegraphics*[width=\textwidth,clip]{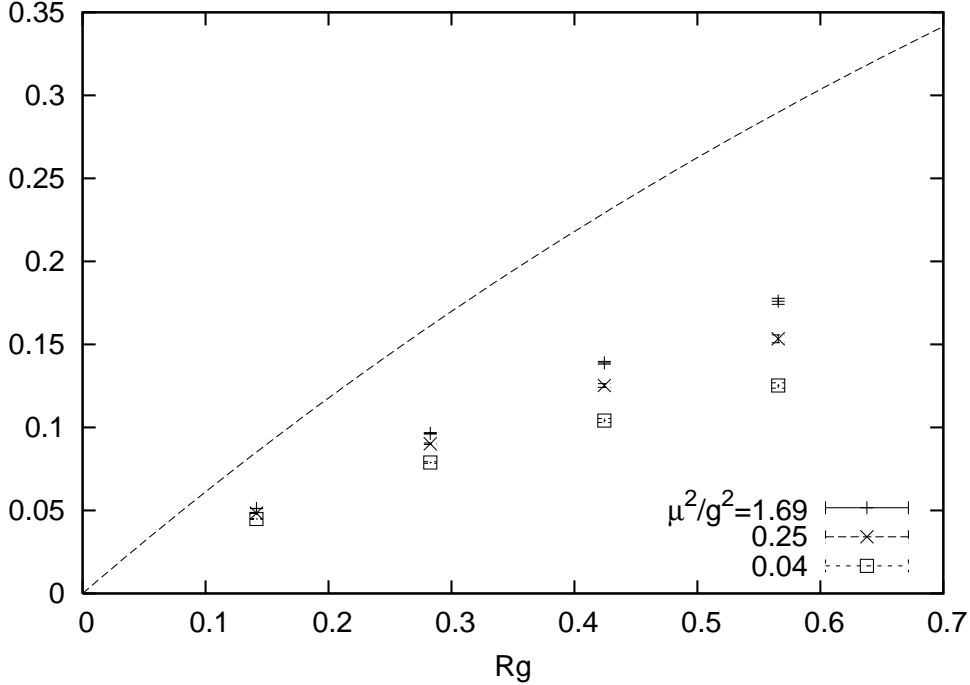}
\end{center}
\caption{$V(R)/g$ determined by linear $\chi^2$-fit described in the text. We
used the configuration sets in~Table~\ref{table:3} and, for $\mu^2/g^2=0.25$,
set~VI of~Table~\ref{table:2}. Eq.~(\ref{twelve}) with $N_c=2$ is plotted by
the broken line.}
\label{fig:5}
\end{figure}

It is also of interest to see how the behavior in Fig.~\ref{fig:4} changes as a
function of the scalar mass. For example, in the limit $\mu^2/g^2\to\infty$,
the scalar fields will completely decouple\footnote{A unique UV divergent
diagram that contains scalar loops is a one-loop scalar self-energy. This
contributes to simply shift the tree-level mass $\mu^2$
by~$\sim g^2\ln\{\mu^2/\Lambda^2\}$, where $\Lambda$ is the UV cutoff, and does
not affect a complete decoupling of the scalar fields in the
limit~$\mu^2/g^2\to\infty$.} and our system would become the two-dimensional
$SU(2)$ QCD with an adjoint massless fermion. On the other hand, the
limit $\mu^2/g^2\to0$ would provide a possible definition of the
two-dimensional $\mathcal{N}=(2,2)$ SYM. It is believed that the screening
occurs in both theories, as already noted. To have a rough idea on this issue,
we carried out a preliminary experiment by using sets of configurations listed
in~Table~\ref{table:3}.
\begin{table}
\caption{Sets of uncorrelated configurations used for Fig.~\ref{fig:5}.}
\label{table:3}
\begin{center}
\begin{tabular}{ccccc}
\hline\hline
$\mu^2/g^2$ & lattice size & $ag$ & $\beta g\times Lg$ &
number of configurations\\
\hline
$1.69$ & $20\times10$ & 0.1414 & $2.828\times1.1414$ & 800 \\
$0.04$ & $20\times10$ & 0.1414 & $2.828\times1.1414$ & 800 \\
\hline\hline
\end{tabular}
\end{center}
\end{table}
The results are summarized in~Fig.~\ref{fig:5}.
For both $\mu^2/g^2=1.69$ and $\mu^2/g^2=0.04$, we still see a linear
potential, although the string tension appears somewhat smaller for smaller
$\mu^2/g^2$.\footnote{Consider the case $\mu^2/g^2=0.04$ in~Fig~\ref{fig:5}.
For this case, the Compton wavelength of a (free) scalar particle is
$1/\mu=5.0/g$ and this is several times longer than the physical lattice size.
Thus in this case the scalar field could effectively be regarded as massless
and the points~$\square$ might be regarded as those for the two-dimensional
$\mathcal{N}=(2,2)$ SYM.}
In the figure, just for reference, we also plotted the function
\begin{equation}
   V(R)=\sqrt{\frac{N_c}{\pi}}g
   \left(1-\exp\left\{-\sqrt{\frac{N_c}{\pi}}gR\right\}\right)
\label{twelve}
\end{equation}
with $N_c=2$, that is given by a semi-classical analysis of a bosonized version
of the two-dimensional massless QCD~\cite{Gross:1995bp}. Strictly speaking, the
overall proportionality constant is not determined by this analysis and we have
chosen it as above without any special reason.

Fig.~\ref{fig:5} is simply a result with a single lattice spacing and a single
lattice size. It is thus not clear what is the real behavior in the continuum
and the large volume limits. Our result is still preliminary and a further
detailed numerical study is needed.

\section{Conclusion}

In this Letter, we illustrated some numerical use of the lattice
formulation~\cite{Sugino:2004qd} of the two-dimensional $\mathcal{N}=(2,2)$ SYM
with a (small) supersymmetry breaking scalar mass. Two physical problems were
considered. For the first one~(Sec.~\ref{sec:2}), our Monte Carlo result fairly
reproduced theoretical prediction on the basis of global symmetries and
(approximate) supersymmetry in the continuum theory. For the second
one~(Sec.~\ref{sec:3}), our result for the static potential~$V(R)$ did not
exhibit the screening behavior that theoretically anticipated. However, since
our result of~$V(R)$ was limited for $Rg\lesssim1$, it is desirable to carry
out a further systematic study by using finer and larger lattices.

We would like to thank Koji Hashimoto and Daisuke Kadoh for helpful
discussions. We thank also the authors of the
FermiQCD/MDP~\cite{DiPierro:2000bd,DiPierro:2005qx} and of a Remez algorithm
code~\cite{Remez} for making their codes available. Our numerical results were
obtained using the RIKEN Super Combined Cluster (RSCC). I.~K.\ is supported by
the Special Postdoctoral Researchers Program at RIKEN. The work of H.~S.\ is
supported in part by a Grant-in-Aid for Scientific Research, 18540305.



\end{document}